# Enhanced critical-current in P-doped BaFe$_2$As$_2$ thin films on metal substrates arising from poorly aligned grain boundaries

Hikaru Sato[1], Hidenori Hiramatsu[1,2,*], Toshio Kamiya[1,2], and Hideo Hosono[1,2]

[1] *Laboratory for Materials and Structures, Institute of Innovative Research, Tokyo Institute of Technology, Mailbox R3-1, 4259 Nagatsuta-cho, Midori-ku, Yokohama 226-8503, Japan*

[2] *Materials Research Center for Element Strategy, Tokyo Institute of Technology, Mailbox SE-6, 4259 Nagatsuta-cho, Midori-ku, Yokohama 226-8503, Japan*

*Correspondence and requests for materials should be addressed to H.Hi. (email: h-hirama@lucid.msl.titech.ac.jp)

Abstract

Thin films of the iron-based superconductor BaFe$_2$(As$_{1-x}$P$_x$)$_2$ (Ba122:P) were fabricated on polycrystalline metal-tape substrates with two kinds of in-plane grain boundary alignments (well aligned (4°) and poorly aligned (8°)) by pulsed laser deposition. The poorly aligned substrate is not applicable to cuprate-coated conductors because the in-plane alignment >4° results in exponential decay of the critical current density ($J_c$). The Ba122:P film exhibited higher $J_c$ at 4 K when grown on the poorly aligned substrate than on the well-aligned substrate even though the crystallinity was poorer. It was revealed that the misorientation angles of the poorly aligned samples were less than 6°, which are less than the critical angle of an iron-based superconductor, cobalt-doped BaFe$_2$As$_2$ (~9°), and the observed strong pinning in the Ba122:P is attributed to the high-density grain boundaries with the misorientation angles smaller than the critical angle. This result reveals a distinct advantage over cuprate-coated conductors because well-aligned metal-tape substrates are not necessary for practical applications of the iron-based superconductors.

Superconductors play an important role in industry and medical instruments, such as high power magnets and magnetic resonance imaging scanners, as well as in condensed matter physics. The refrigeration cost (e.g., the cost of liquid helium) is the most important issue for superconductor applications. However, the discovery of high critical temperature ($T_c$) superconductivity in a copper-based oxide (cuprate) in 1986[1] provides the possibility to overcome this issue. To apply superconductors to practical devices such as high power magnets, flexible superconducting wires are critically important, but it is difficult for oxides to make them because of their brittleness. Moreover, high-$T_c$ cuprates have another serious issue originating from their weak-link behaviour at grain boundaries due to their unconventional pairing symmetry (*d*-wave), short coherence length, and low carrier concentration[2]. Their critical current densities $J_c$ drastically decrease at grain boundaries if adjacent crystalline grains have in-plane misorientation angles $\geq 4°$[3]. Therefore, the crystallographic orientations of the grains in cuprate superconducting wires/tapes/coated conductors must be very well aligned to maintain high superconductivity performance. Thus, the in-plane misorientation angles should be as small as possible to overcome the $J_c$ issue. For Bi–Sr–Ca–Cu–O flat tapes / round wires (called Bi2223 and Bi2212)[4], rolling processes and post-rolling heat treatment are necessary to exactly align their crystallographic *ab*-plane direction to the current direction as well as to achieve small grain boundary angles[5]. For coated conductors using $YBa_2Cu_3O_{7-\delta}$[6] and other rare-earth-based cuprates (REBCO), two types of techniques are used to achieve highly in-plane oriented practical metal-tape substrates. One technique is to deposit a biaxially textured buffer-layer on a non-oriented metal-tape by the ion beam assisted deposition (IBAD) method[7,8]. The other is to use a highly oriented Ni–W alloy metal-tape substrate fabricated by the rolling-assisted biaxially textured substrate (RABiTS) method[9,10]. In the typical IBAD process, a planarizing bed layer, such as amorphous (*a*-)$Y_2O_3$ or *a*-$Ga_2Zr_2O_7$ (*a*-GZO), is first deposited on a non-oriented polycrystalline Ni-based Hastelloy substrate. A biaxially textured oxide layer, such as MgO, GZO, or yttria-stabilized zirconia, is then formed by IBAD, followed by sequential deposition of homoepitaxial MgO, $LaMnO_3$ (LMO), and/or $CeO_2$ layers to achieve extremely high in-plane orientation $\Delta\phi_{CeO2}$ or $\Delta\phi_{MgO} < 4°$ in full width at half maximum (FWHM) (see Fig. 1)[11–15]. In contrast, RABiTS uses an oriented Ni–W alloy as the starting metal-tape substrate. In this case, some oxide buffer layers, such as $CeO_2$, are also used[16]. To produce high-performance REBCO coated conductors, IBAD is the most powerful method because it produces a top oxide-layer with better in-plane alignment than that produced by RABiTS. However, IBAD needs a number of deposition steps to produce the highly oriented top layer, leading to high cost and low throughput. Thus, simpler metal-tape substrates with less in-plane orientation alignment are advantageous in terms of production cost.

In 2008, a new type of high-$T_c$ superconductor, iron-based superconductors, was discovered[17]. These materials have attractive properties for superconducting wires/coated conductors because they have extremely high upper critical magnetic field ($H_{c2}$)[18], low anisotropy factor $\gamma = (H_{c2}//ab)/(H_{c2}//c) = 1–5$[19], and high $T_c$ (maximum ~55 K)[20]. The most important property of iron-based superconductors is their advantageous grain boundary nature. Katase *et al.*[21] clarified that an iron-based superconductor, Ba($Fe_{1-x}Co_x$)$_2$$As_2$ (Ba122:Co), has the doubly larger critical grain boundary misorientation angle for $J_c$ ($\theta_c = ~9°$) than cuprates ($\theta_c = ~4°$)[3]. This advantage enables low-cost fabrication processes for superconducting wires/coated conductors using less aligned grain orientations. Therefore, iron-based superconductors are expected to be candidates for a new type of superconducting wires/coated conductors.

Concerning thin films of iron-based superconductors, Ba122:Co epitaxial films have been extensively investigated[22,23] because of the low vapor pressure of the Co dopant and the easy growth of high-quality films[24–31]. $BaFe_2(As_{1-x}P_x)_2$ (Ba122:P) exhibits a higher maximum $T_c$ (~31 K)[32] than Ba122:Co (~22 K)[33]. Therefore, Ba122:P thin films have also been extensively investigated[34–39]. We recently reported that Ba122:P epitaxial films on MgO single crystals exhibit high $J_c$ values[38], where *c*-axis-oriented vortex pinning centers were intentionally introduced by optimizing the growth rate (2.2 Å/s) and temperature (1050 °C). Consequently, $J_c$ drastically increased and exceeded 1 MA/cm$^2$ at 9 T along the *ab*-plane at 4 K with highly isotropic properties against external magnetic fields, which is the highest $J_c$ value among the iron-based superconductor thin films reported so far.

In this study, we fabricated Ba122:P thin films on metal-tape substrates by pulsed laser deposition (PLD) using the optimized conditions reported in ref. 38. Figure 1(b) shows a schematic diagram of the stacking structure of the IBAD-MgO substrate used in this study. Two types of metal-tape substrates fabricated by the IBAD method were used[12,14]. One was well-aligned IBAD-MgO ($\Delta\phi_{MgO}$ = 4°) and the other was poorly aligned IBAD-MgO ($\Delta\phi_{MgO}$ = 8°). Compared with typical IBAD metal-tapes used for REBCO-coated conductors (Fig. 1(a)), the in-plane alignment of poorly aligned IBAD-MgO substrates ($\Delta\phi_{MgO}$ = 8°) is too large and they cannot be used for REBCO-coated conductors. Unexpectedly, the Ba122:P film on poorly aligned IBAD-MgO exhibited higher $J_c$ with stronger vortex pinning properties than the Ba122:P film on well-aligned IBAD-MgO, demonstrating its potential for iron-based superconductors. The origin of the strong pinning properties is discussed based on microstructure analysis and related to the dislocation structures.

**Results and discussion**

*Structure and electronic transport properties*

Figure 2(a) shows the out-of-plane X-ray diffraction (XRD) patterns of Ba122:P films on the two types of IBAD-MgO substrate. We confirmed the strong *c*-axis orientation of the Ba122:P films on both the IBAD-MgO substrates, while small non-oriented crystallite domains (determined from weak 110 and 112 diffractions) were detected only in the film on poorly aligned IBAD-MgO. From the peaks at $2\theta = \sim29$ and 33°, the planarizing *a*-$Y_2O_3$ bed layers in the IBAD-MgO substrates slightly crystallized. Because the Ba122:P films exhibit four-fold symmetric peaks in the $\phi$ scans of asymmetric 103 diffraction owing to the tetragonal symmetry (Fig. 2(b)), the Ba122:P films pseudo-epitaxially grew on both the IBAD-MgO substrates with the orientation relation of Ba122:P [001] ∥ IBAD-MgO [001] out-of-plane and Ba122:P [100] ∥ IBAD-MgO [100] in-plane without extra in-plane rotational domains. Concerning the crystallite quality of the two types of Ba122:P film, the FWHM values of out-of-plane diffraction (Fig. 2(c)) are almost the same ($\Delta\omega_{Ba122:P} = 1.3°$), whereas those of in-plane diffraction (Fig. 2(d)) are greatly different. The FWHM of the $\phi$ scan of 200 in-plane diffraction peak of the Ba122:P film on poorly aligned IBAD-MgO ($\Delta\phi_{Ba122:P} = 8.0°$) is much larger than that for the Ba122:P film on well-aligned IBAD-MgO ($\Delta\phi_{Ba122:P} = 2.7°$), which is mainly because of the different in-plane alignment of the IBAD-MgO metal-tape substrates. A similar result is also observed in the $\phi$ scans of asymmetric 103 diffraction (Fig. 2(b)). This difference would lead to growth of some non-oriented crystallite domains (110 and 112 diffractions) observed in the out-of-plane XRD patterns of the film on poorly aligned IBAD-MgO.

In the case of growth on MgO single-crystal substrates[37,38], in-plane tensile strain is introduced in Ba122:P films mainly because of large in-plane lattice mismatch between Ba122:P and MgO (~+7%). Similarly, the in-plane lattice parameters of the Ba122:P films on both the IBAD-MgO substrates are larger than that of bulk Ba122:P ($a$ = 0.392 nm)[32]. The $a$-axis lengths of Ba122:P on well-aligned and poorly aligned IBAD-MgO are similar ($a$ = 0.3935 and 0.3948 nm, respectively). The $c$-axis lengths of the Ba122:P films on both well-aligned and poorly aligned IBAD-MgO are also almost the same (1.2824 and 1.2840 nm, respectively). The chemical composition ratios of P to As ($x$ = P/(As+P)) of the Ba122:P films are $x$ = 0.283 ± 0.06 for the well-aligned sample and $x$ = 0.261 ± 0.06 for the poorly aligned sample, which are both less than that for the optimal P concentration of bulk Ba122:P of $x$ = ~0.33[32]. This is ascribed to the slightly underdoped region of the phase diagram irrespective of the use of the higher P concentration PLD target ($x$ ≈ 0.35). Similarly poor transferability of P from a target to a film was observed in ref. 38 because of the relatively high vapor pressure of P. Figure 2(e) shows the temperature dependence of the resistivity of Ba122:P films on both the IBAD-MgO substrates (200 nm thick Ba122:P for the well-aligned and 150 nm thick for the poorly aligned samples; Note that these samples are different from those observed by STEM in Fig. 4, which will be shown later on.). $T_c$ of the well-aligned sample ($T_c^{onset}$ = 26 K, $T_c^{zero}$ = 23 K) is slightly higher than that of the poorly aligned sample ($T_c^{onset}$ = 23 K, $T_c^{zero}$ = 19 K). The main reason for the higher $T_c^{onset}$ in the well-aligned sample is the slight difference in the P concentration (~2%), that is, the P concentration of the well-aligned sample is closer to the optimum P concentration than that of the poorly aligned sample.

*Transport $J_c$ values*

As mentioned above, the in-plane crystallinity of the Ba122:P film on poorly aligned IBAD-MgO is much poorer than that on well-aligned IBAD-MgO. However, the transport $J_c$ values of the poorer film are better in terms of coated conductor applications. Figure 3(a) shows the transport $J_c$ of Ba122:P films (the thicknesses of the well-aligned and the poorly aligned Ba122:P films are 200 and 150 nm, respectively. Note that these samples are different from those observed by STEM in Fig. 4, which will be shown later on.) on both the IBAD-MgO substrates at 4 and 12 K as a function of the external magnetic field (*H*). The self-field $J_c$ values at each temperature [1.02 (at 4 K) and 0.37 MA/cm$^2$ (at 12 K) for the well-aligned sample, and 1.19 (at 4 K) and 0.39 MA/cm$^2$ (at 12 K) for the poorly aligned sample] are almost the same for both the samples. These values are comparable with those of a Ba122:Co (1.6 MA/cm$^2$ at 4 K [40]) and a Ba122:Co/Fe (2 MA/cm$^2$ at 4 K [41]), and higher than those of a Ba122:Co/Fe (0.1 MA/cm$^2$ at 8 K [42]) and a NdFeAsO coated conductors (0.071 MA/cm$^2$ at 5 K [43]). It should be noted that the $J_c$ values of the Ba122:P film on poorly aligned IBAD-MgO at 4 K under magnetic fields up to 9 T are higher than those of the film on well-aligned IBAD-MgO for the same temperature and *H* direction. Although the $T_c$ and crystallinity are poorer than those of the film on well-aligned IBAD-MgO, this result indicates that the poorly aligned Ba122:P film has a stronger vortex pinning property. The maximum $J_c$ at 9 T (0.16 MA/cm$^2$ at 4 K) is comparable with those of a Ba122:Co (0.16 MA/cm$^2$ at 4 K [40]) and a Ba122:Co/Fe (0.1 MA/cm$^2$ at 4 K [41]) much higher than those of a Ba122:Co/Fe (3 kA/cm$^2$ at 8 K [42]) and NdFeAsO coated conductors [43], but slightly lower than that of an FeSe coated conductor (0.7 MA/cm$^2$ at 4

K [44]).

To investigate the vortex pinning properties, the magnetic-field angular ($\theta_H$) dependence of the transport $J_c$ was measured. Figures 3(c) and (d) show the $\theta_H$ dependence of $J_c$ at 4 and 12 K, respectively. $J_c$ peaks are observed both at $\theta_H = 90°$ (*H* // *ab*) and $\theta_H = 0°$ (*H* // *c*), leading to isotropic properties similar to those of the films on MgO single crystals[38]. The former corresponds to the intrinsic property (i.e., electronic anisotropy) and/or some effects from the existence of pinning centers parallel to the *ab* plane (e.g., stacking faults), whereas the latter corresponds to the existence of pinning centers along the *c* axis. For the Ba122:P film on poorly aligned IBAD-MgO, the $J_c$ values at 4 K and the peaks for *H* // *c* remain high even at 9 T compared with those of the Ba122:P film on well-aligned IBAD-MgO. These results indicate that stronger and/or larger numbers of vortex pinning centers along the *c* axis are introduced in the film on the poorly aligned IBAD-MgO substrate. As shown in Figure 2(d), the Ba122:P film on poorly aligned IBAD-MgO has poorer in-plane crystallinity ($\Delta\phi_{Ba122:P} = 8.0°$) but the $J_c$ performance is better than that of the Ba122:P film on well-aligned IBAD-MgO. This result indicates that introduction of pinning centers owing to the poorer crystallinity enhances the $J_c$ performance. On the other hand, the relationship between $J_c$ of the poorly aligned sample and the well-aligned one becomes opposite between 4 K and 12 K. $J_c$ at 12 K are almost the same for both the samples up to 5 T, but the relation changed in the higher *H* region, and the $J_c$ of the poorly aligned sample becomes smaller than that of the well-aligned one. This rapid decay of $J_c$ of the poorly aligned sample in the high *H* region at 12 K would be related to its low $T_c^{zero}$. These results indicate that the pinning centers consist of small-sized defects such as dislocations and are not large-sized impurity phases, which will be discussed later on.

*Microstructure analysis*

To clarify the origin of the better $J_c$ performance and strong vortex pinning in the film on poorly aligned IBAD-MgO, we observed the microstructures by cross-sectional scanning transmission electron microscopy (STEM). Figures 4(a) and (b) show bright-field STEM images of Ba122:P on well-aligned and poorly aligned IBAD-MgO, respectively. Both images show a clear stacked structure composed of the Ba122:P film, epitaxial MgO, IBAD-MgO, $Y_2O_3$, and Hastelloy substrate, where deterioration of each layer was not detected irrespective of the high temperature growth (1050 °C). In contrast to a Ba122:Co film fabricated by low-power KrF excimer laser PLD[45], horizontal defects such as stacking faults are not observed in the present films, indicating that the peaks along the *ab* planes in the $J_c$–$\theta_H$ plots mainly originate from the intrinsic electronic anisotropy of Ba122:P. In addition, $J_c$ values for *H* // *ab* at 4 K of the Ba122:P film on poorly aligned IBAD-MgO are higher than those of Ba122:P film on well-aligned aligned IBAD-MgO (see Fig. 3(c)). Because horizontal defects are not observed, one possible origin of this higher $J_c$ along the *ab* plane would be point-like defects, which work as more isotropic pinning centers as reported in proton-irradiated Ba122:Co[46]. Moreover, vertical defects are observed in both Ba122:P films, which is consistent with the result that $J_c$ peaks are observed along the *c* axis in the $J_c$–$\theta_H$ plots. However, the vertical defects in the well-aligned sample are tilted largely compared to those in the poorly aligned one, and the density of the vertical defects in the poorly aligned sample is clearly higher than that in the well-aligned one. The larger tilt angle for the well aligned sample is consistent with the wider angular distribution along the *c*-axis of the $J_c$–$\theta_H$ plots for the well-aligned sample observed in Fig. 3(c). Considering the higher $J_c$ performance of the Ba122:P film on the poorly aligned IBAD-MgO substrate, as observed in Fig. 3(c), these vertical defects would act as effective pining centres and their density dominates the $J_c$ value under external magnetic fields.

To clarify the type of vertical defects, the chemical compositions of the vertical defects were determined. Figure 4(c) shows the results of line scans for each element by energy dispersion X-ray (EDX) spectroscopy. As shown in the lower panel of Fig. 4(c), the chemical composition of a vertical defect is the same as that in the bulk region of the Ba122:P film, indicating that the introduced vertical defects should be dislocations or domain boundaries rather than an impurity phase. This result is the same as Ba122:P epitaxial films on single-crystal MgO[38] but different from the case of an oxygen-diffused Ba122:Co film on SrTiO$_3$ (in this case, oxide nanorods act as strong pinning centres)[47]. To investigate oxygen diffusion from IBAD-MgO to the Ba122:P film in more detail, high-angle annular dark-field (HAADF)-STEM was performed over a heterointerface between IBAD-MgO and Ba122:P (Fig. 4(d)) because HAADF-STEM emphasizes the contrast of the atomic numbers of the constituent elements. Part of the heterointerface shows a different contrast in the HAADF-STEM image (the dashed red closed area in the left panel of Fig. 4(d)), indicating that the chemical composition is slightly different from the bulk region of the Ba122:P film. As observed in the result of an EDX line scan over the heterointerface (right panel of Fig. 4(d)), slight oxygen diffusion from the film into IBAD-MgO of ~20 nm in depth is detected. Because the out-diffused region in the film is limited to the vicinity of the interface, it does not act as dominant vortex pinning centres. In the case of a film on single-crystal MgO[38], such an deep oxygen-diffusion region was not observed. Thus, formation of the oxygen diffusion region is enhanced by the surface structure of IBAD-MgO composed of small grains[48].

According to the cross-sectional STEM images and chemical composition analysis, the density of the vertical defects (not impurities) is the dominant factor determining the $J_c$ value under a magnetic field. To determine the defect structure, plane-view STEM was performed. Figure 5 shows plane-view bright-field STEM images of the Ba122:P film on two types of IBAD metal-tape substrates with $\Delta\phi_{MgO} = 4°$ (well-aligned, (b) and (d)) and 8° (poorly aligned, (c) and (e)). As shown in Fig. 4(b), the Ba122:P film on the poorly aligned IBAD-MgO substrate is composed of high-density vertical grain boundaries. It is known that grain boundaries act as pinning centers[49,50] but they also lead to decay of $J_c$ if their misorientation angle is larger than the critical angle for $J_c$ (e.g., $\theta_c = \sim9°$ for an iron pnictide superconductor, Ba122:Co [21]). Comparing both the plane-view images, the grain sizes of the well-aligned sample are larger than those of the poorly aligned sample although the size distribution is relatively large (the lateral grain sizes are 120 – 230 nm for the well-aligned sample and 80 – 150 nm for the poorly aligned sample). These results indicate that the grain boundary density of the poorly aligned sample is higher than that of the well-aligned sample, which is consistent with the stronger vortex pinning properties of the poorly aligned sample as observed in Fig. 3(c). To investigate the depth structure of the boundaries, we obtained plane-view STEM images by tilting the film's surface with respect to the incident electron beam direction, as shown in the bottom right panels of Figs. 5(d) and (e). As shown by the arrows in the enlarged image (the upper right panel of Figs. 5(d) and (e)), the grain boundaries have arrays of dislocations almost normal to the grain boundary planes for both samples. This type of defect structure (i.e., arrays of dislocations) indicates that these boundaries are low misorientation-angle grain boundaries. The distances between adjacent dislocations are 9.4 – 18 nm for the well aligned sample (Fig. 5(d)) and 3.8–6.3 nm for the poorly aligned sample (Fig. 5(e)), which respectively represent misorientation angles of 1.3–2.4° and 3.6–5.9° calculated by the dislocation space ($D$) – misorientation angle at a grain boundary ($\theta_{GB}$) relation $D = (|b|/2)/\sin(\theta_{GB}/2)$, where $|b|$ is the norm of the corresponding Burgers vector[3]. Although these estimated $\theta_{GB}$ for both samples are slightly less than the FWHM values of the in-plane XRD rocking curves (see Fig. 2(d)), both FWHM and $\theta_{GB}$ are clearly less than $\theta_c = \sim 9°$ for an iron pnictide superconductor[21], and their relative relationship is the same for both the samples. Therefore, the Ba122:P film on poorly aligned IBAD-MgO has high density grain boundaries with small misorientation angles, so they do not decrease its $J_c$ performance and effectively act as vortex pinning centres. Using the advantageous grain boundary properties of iron-based superconductors, lower quality and lower cost metal-tape substrates would improve the vortex pinning properties by naturally forming grain boundaries that do not decrease the $J_c$ performance.

**Summary**

Ba122:P films deposited on IBAD-MgO substrates show promising properties for future practical applications. The film on the poorly aligned IBAD-MgO substrate exhibited stronger vortex pinning properties and higher $J_c$ values than that on a well-aligned IBAD-MgO substrate. According to microstructure analysis, a higher density of vertical defects, which act as stronger pinning centres along the *c* axis, was introduced in the film on the poorly aligned IBAD-MgO substrate than in the film on the well-aligned IBAD-MgO substrate. These vertical defects correspond to grain boundaries that have smaller misorientation angles than the critical angle for $J_c$ of iron-based superconductors (9°). These results demonstrate the clear advantage of Ba122:P-coated conductors over cuprate-coated conductors because high-quality (i.e., highly textured by stacking many layers) metal-tape substrates are not needed to produce high performance coated conductors if the misorientation angles are successfully controlled to be less than the critical angle, although the critical current of the Ba122:P coated conductors are still to be improved by increasing their thickness to achieve a critical current comparable to REBCO.

**Methods**

**Thin-film growth**

Ba122:P films with thicknesses of 150–200 nm were grown on IBAD-MgO metal-tape substrates (10 mm × 10 mm) supplied by iBeam Materials, Inc. (NM, USA) (see Fig. 1(b) for their structure)[12,14]. As it was difficult to measure Ba122:P film thickness with a stylus profiler due mainly to the curvature of the IBAD-MgO metal-tape, film thickness was determined by cross-sectional field-effect scanning electron microscope (FE-SEM) observation where the sample was cut out by focused ion-beam (FIB). The detailed procedure to determine the thickness is summarized in Supplementary Fig. S1. IBAD-MgO substrates with $\Delta\phi_{MgO}$ = 4° and 8°, which were controlled by the thickness of the top MgO layer, were employed; i.e., the nominal MgO thicknesses of 120 and 65 nm corresponded to $\Delta\phi_{MgO}$ = 4° and 8°, respectively. Polycrystalline $BaFe_2(As_{0.65}P_{0.35})_2$ disks ($x$ = 0.35) were used as the PLD targets. The other growth parameters (e.g., the growth temperature and rate of 1050 °C and 2.2 Å/s, respectively) were the same as those optimized for Ba122:P films on single-crystal substrates[38].

**Characterization**

$\omega$-coupled $2\theta$ scan XRD measurements were performed to determine the crystalline phases. Asymmetric 103 diffraction of the Ba122:P films was measured to confirm the in-plane crystallographic symmetry. The crystallinity of the films was characterized on the basis of the FWHM of the out-of-plane 004 ($\Delta\omega$) and in-plane 200 rocking curves ($\Delta\phi$). The relationship of these axes for XRD can be found in ref. 51. The chemical compositions were determined with an electron-probe microanalyzer, where the acceleration voltage of the electron beam was adjusted while monitoring the Ni K$\alpha$ spectrum to avoid the matrix effect from the Ni-containing Hastelloy metal-tapes (see Supplementary Fig. S2). For quantitative analyses, we employed the atomic number, absorption, fluorescence (ZAF) correction method using the following standard samples; $BaTiO_3$ for Ba, Fe for Fe, LaAs for As, and InP for P. Because the Ba:Fe atomic ratio was confirmed to be stoichiometric (i.e., Ba:Fe = 1:2), the chemical composition ratios of P to As of the Ba122:P films were calculated using the atomic ratio $x$ = P/(As+P).

To investigate the electrical transport properties, the films were patterned into microbridges by photolithography and Ar milling with the length of 300 μm and the widths of 10 μm for the well-aligned sample and 20 μm for the poorly aligned sample. The temperature dependence of the electrical resistivity was measured by the four-probe method with a physical property measurement system. The transport $J_c$ was determined from voltage–current curves with the criterion of 1 μV/cm under external magnetic fields ($H$) up to 9 T in the maximum Lorentz force configuration ($J \perp H$). The angle ($\theta_H$) of the applied $H$ was varied from −30 to 120°. $\theta_H$ = 0 and 90° correspond to the configurations of *H* // *c* axis (i.e., normal to the film surface) and *H* // *ab* plane, respectively. The relationship between *J*, *H*, and $\theta_H$ are shown in Fig. 3(b).

Microstructure analysis was performed with a STEM system. To observe the plane-view and cross-sectional images, all of the samples were prepared by the FIB microsampling technique and then directly transferred to the STEM system under high vacuum without exposure to air to avoid contamination and oxidization of the samples. The chemical compositions near defects and heterointerfaces were evaluated by EDX line scans with a spatial resolution of ~1 nm.

**Acknowledgements**

This work was supported by the Ministry of Education, Culture, Sports, Science and Technology (MEXT) through the Element Strategy Initiative to Form Core Research Center. H.Hi. was also supported by the Japan Society for the Promotion of Science (JSPS) through a Grant-in-Aid for Young Scientists (A) (Grant Number 25709058), a JSPS Grant-in-Aid for Scientific Research on Innovative Areas "Nano Informatics" (Grant Number 25106007), and Support for Tokyotech Advanced Research (STAR).

**Author Contributions**

H.Hi. and H.Ho. designed the research. H.S., H.Hi., and T.K. carried out the experiments and the data analyses, and wrote the manuscript with discussion with H.Ho.

**Additional Information**

Supplementary information accompanies this paper at http://www.nature.com/srep

**Competing financial interests**: The authors declare no competing financial interests.

**Figures**

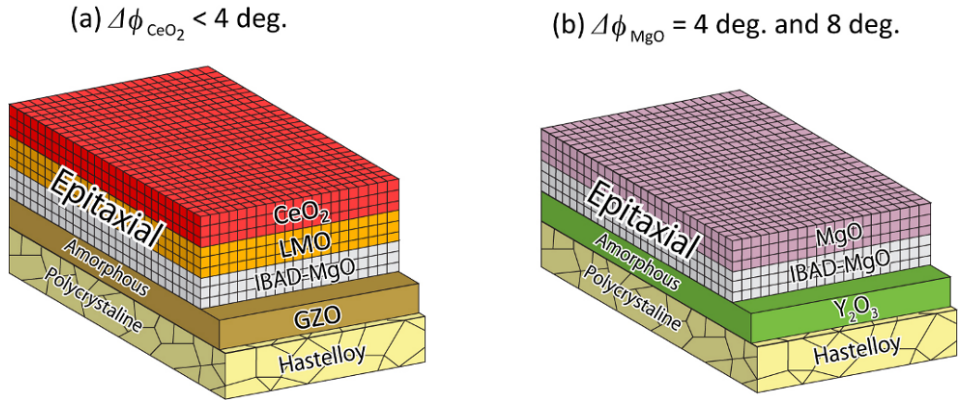

**Figure 1. Schematic diagrams of the structures of flexible metal-tape substrates fabricated by the IBAD method.** The bottom metal-tape substrate is non-oriented Ni-based Hastelloy. (**a**) For high-$T_c$ cuprate (ex. REBCO) superconductors. Amorphous $Ga_2Zr_2O_7$ (*a*-GZO) as a cation diffusion barrier is first formed, followed by sequential deposition of biaxially textured MgO by the IBAD method, $LaMnO_3$ (LMO) by sputtering, and the top $CeO_2$ layer by PLD to achieve extremely high in-plane orientation with $\Delta\phi_{CeO2} < 4°$[13,15]. (**b**) For iron-based superconductors (IBAD-MgO). An *a*-$Y_2O_3$ bed layer is deposited on the metal-tape substrate to planarize the surface, followed by sequential deposition of biaxially textured MgO by the IBAD method and a top homoepitaxial MgO layer[12,14]. In this study, we used two types of IBAD-MgO substrate with in-plane alignment of $\Delta\phi_{MgO}$ = 4 and 8°.

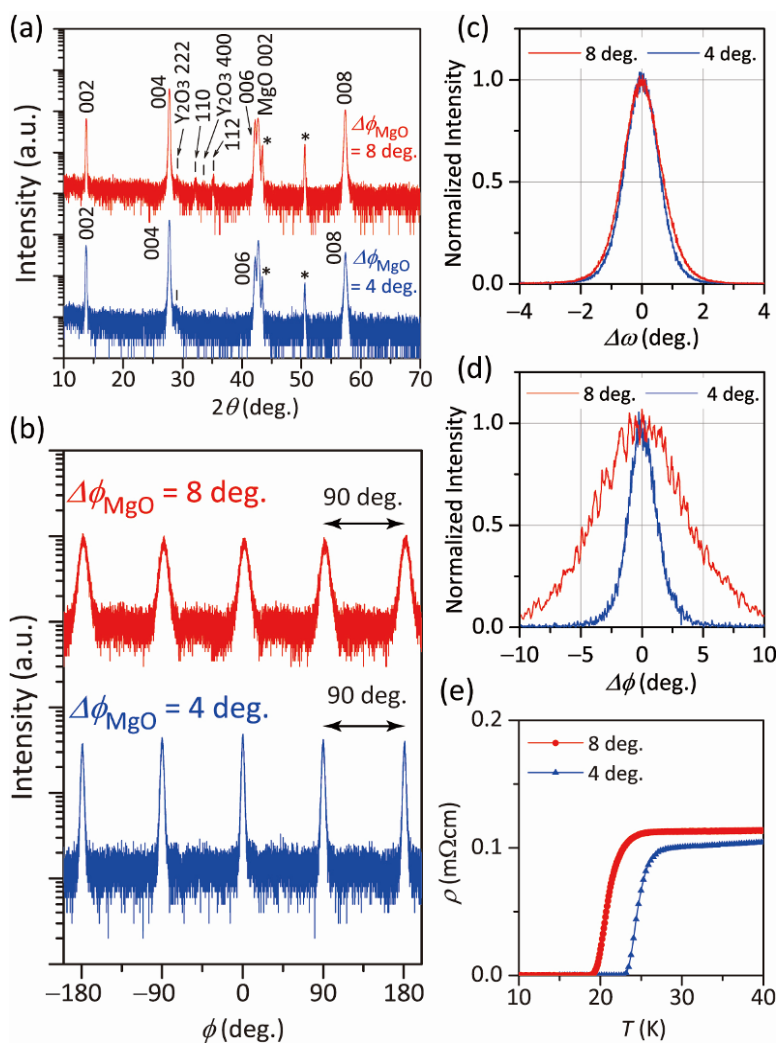

**Figure 2. Structural and electronic transport properties of Ba122:P films on two types of IBAD-MgO metal-tape substrate with Δ$\phi_{MgO}$ = 8° (poorly aligned) and 4° (well aligned).** (**a–d**) XRD patterns. (**a**) $\omega$-coupled $2\theta$ scans for out-of-plane. The asterisks indicate the diffraction peaks from the Hastelloy. (**b**) $\phi$ scans of asymmetric 103 diffraction. (**c**, **d**) Intensity-normalized rocking curves of (**c**) out-of-plane 004 and (**d**) in-plane 200 diffraction. (**e**) Temperature ($T$) dependence of the resistivity ($\rho$).

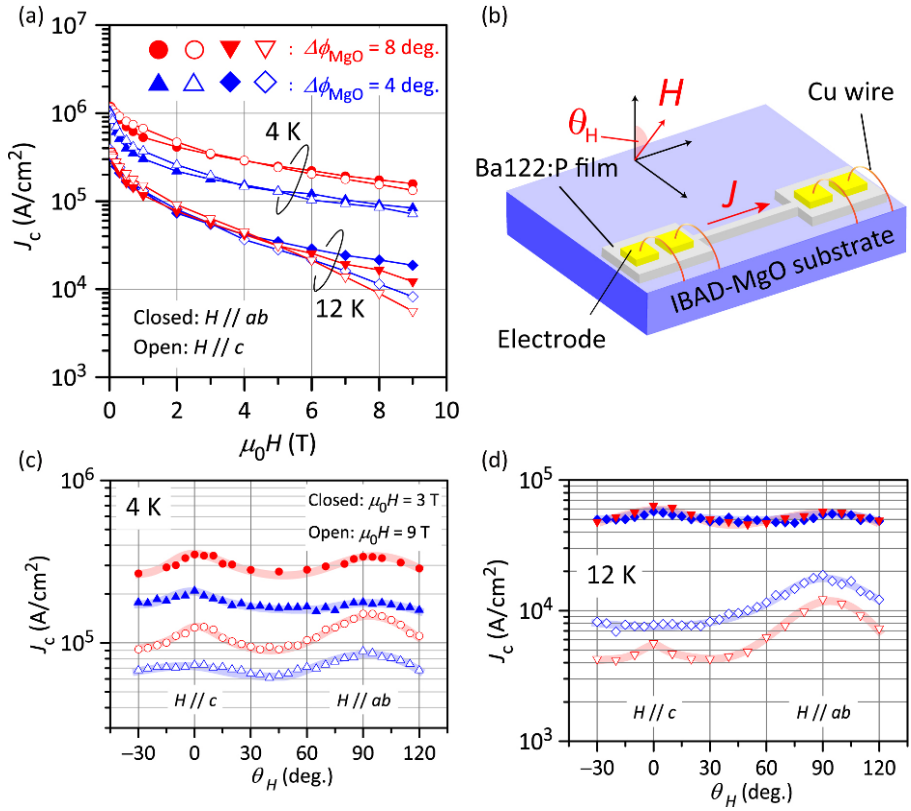

**Figure 3. Critical current density ($J_c$) of Ba122:P films on two types of IBAD metal-tape substrate with $\Delta\phi_{MgO}$ = 8° (poorly aligned, red symbols) and 4° (well aligned, blue symbols).** (**a**) External magnetic field ($H$) dependence of $J_c$ at 4 and 12 K. The closed and open symbols indicate the configurations of $H \parallel ab$ plane and $H \parallel c$ axis of the Ba122:P films, respectively. (**b**) Relationship between current ($J$) and $H$ directions under $J_c$ measurement. We applied $H$ up to 9 T in the maximum Lorentz force configuration (i.e., $J \perp H$). The external magnetic field angle ($\theta_H$) was varied from −30 to 120°. $\theta_H$ = 0 and 90° correspond to the configurations of $H \parallel c$ axis (i.e., normal to the film surface) and $H \parallel ab$ plane, respectively. (**c**, **d**) $\theta_H$ dependence of $J_c$ at (**c**) 4 K and (**d**) 12 K. The closed and open symbols are the data under $\mu_0 H$ = 3 and 9 T, respectively.

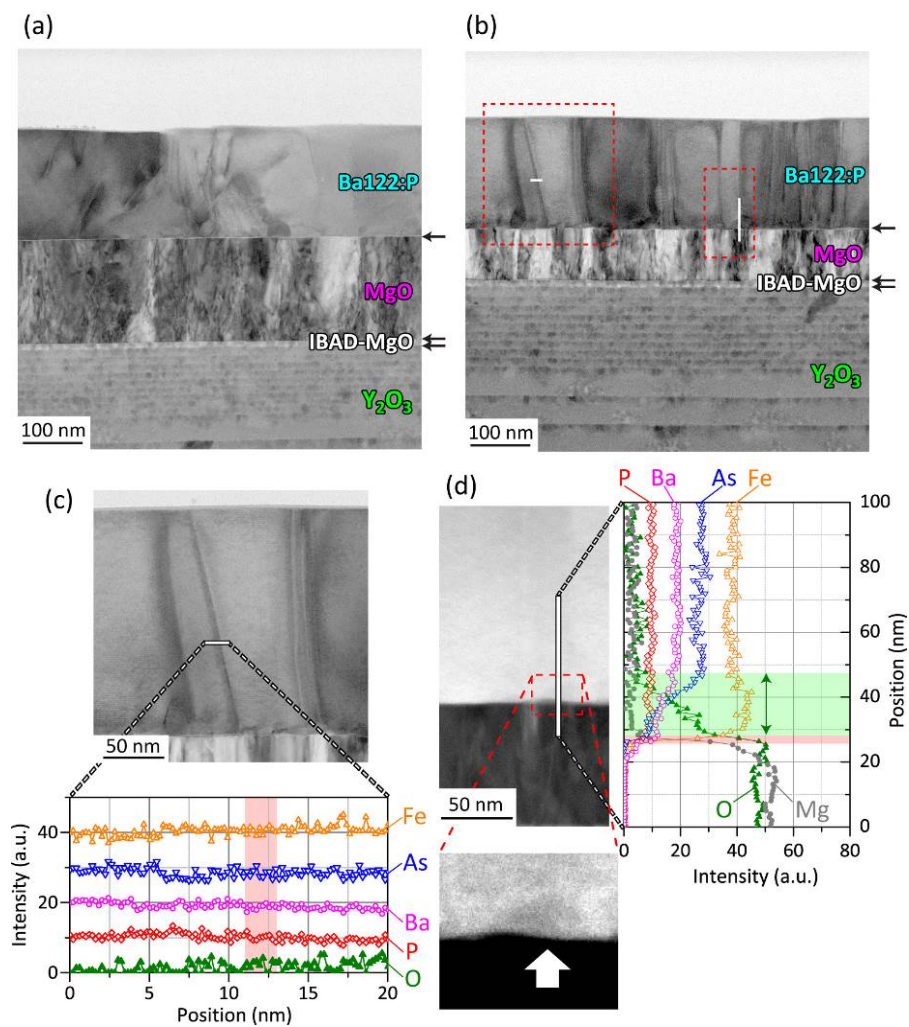

**Figure 4. Cross-sectional microstructure and chemical composition analysis of Ba122:P films on two types of IBAD metal-tape substrate with $\Delta\phi_{MgO}$ = 4° (well aligned) and 8° (poorly aligned).** (**a**, **b**) Bright-field STEM images of Ba122:P films on (**a**) a well-aligned IBAD-MgO substrate with $\Delta\phi_{MgO}$ = 4° and (**b**) a poorly aligned IBAD-MgO substrate with $\Delta\phi_{MgO}$ = 8°. The horizontal arrows indicate heterointerfaces. The top panel in (**c**) shows a magnified image of the left red dashed square in (**b**). The lower panel shows the results of EDX line scans along the horizontal line in the upper panel, where the shaded red area indicates the vertical defect position. The left panel in

(**d**), which corresponds to that indicated in the right red dashed area in (**b**), shows the HAADF-STEM image of the heterointerface between the Ba122:P film and the MgO layer. The dashed red square in the left panel of (**d**) shows a different contrast region at the heterointerface. To show it clearer, the region is enlarged at the bottom where the different contrast region is indicated by the vertical arrow. The right panel in (**d**) shows the results of EDX line scans along the vertical white line in the left panel. The thick horizontal red line and shaded green area indicate the heterointerface and oxygen diffusion area, respectively.

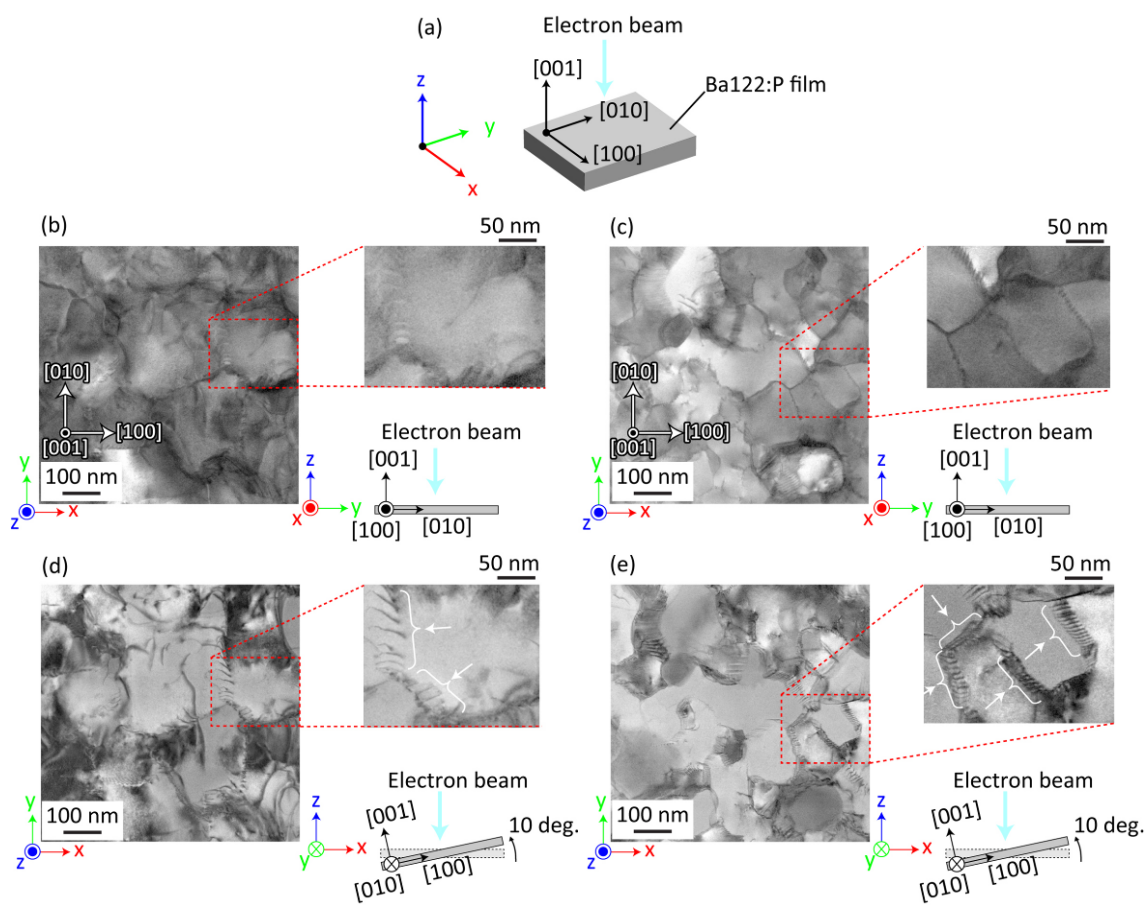

**Figure 5. Plane-view bright-field STEM images of Ba122:P films on two types of IBAD metal-tape substrates with $\Delta\phi_{MgO}$ = 4° (well aligned) and 8° (poorly aligned).** (**a**) Relationship between the film orientation, direction of the incident electron beam, and global axes (*X, Y,* and *Z*). (**b, c**) Typical plane-view images observed by normal electron beam incidence for Ba122:P films on IBAD metal-tape substrates with $\Delta\phi_{MgO}$ = 4° (b) and 8° (c). (**d, e**) Slanted angle images observed by tilting 10° to enhance dislocations. (d) and (e) are the same areas of (b) and (c), respectively (see the bottom right figures for the observation geometries). The top right figures in (b) − (e) are enlarged images of the dashed squares in the left images of (b) − (e), respectively. The arrows in the top right image of (d) and (e) show arrays of dislocations.

Supplementary information for "Enhanced critical-current in P-doped BaFe$_2$As$_2$ thin films on metal substrates arising from poorly aligned grain boundaries"

Hikaru Sato[1], Hidenori Hiramatsu[1,2], Toshio Kamiya[1,2], and Hideo Hosono[1,2]

[1]*Laboratory for Materials and Structures, Institute of Innovative Research, Tokyo Institute of Technology, Japan*

[2]*Materials Research Center for Element Strategy, Tokyo Institute of Technology, Japan*

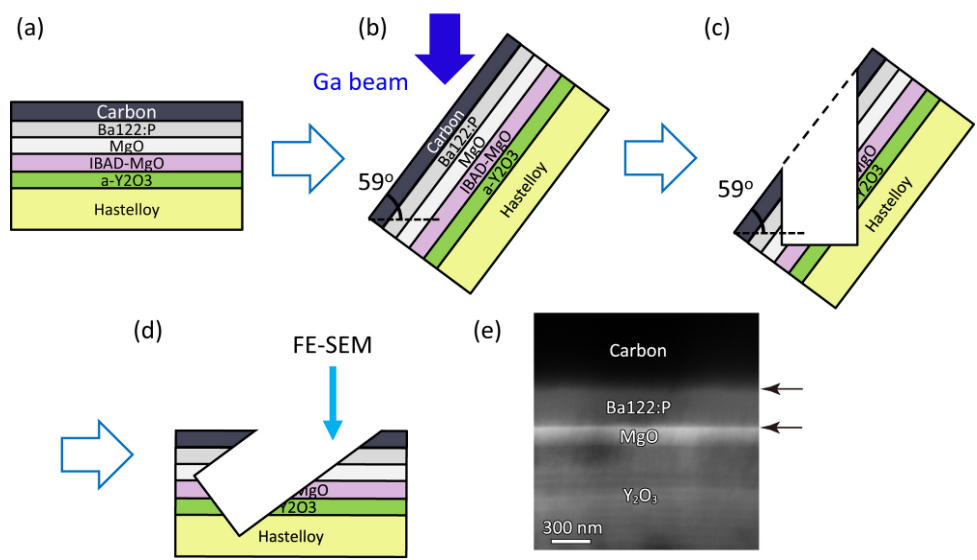

**Figure S1. Determination procedure of the thickness of the Ba122:P film on the IBAD-MgO substrate.** (**a**) Schematic cross-sectional illustration of stacking structure before FIB treatment. (**b**) Configuration of the sample and a gallium-ion beam in FIB system. (**c**) Schematic cross-sectional image after FIB etching processing. (**d**) The sample was transferred to the FE-SEM system to observe the cross section of the FIB-processed position. (**e**) An example of the cross-sectional FE-SEM image of a sample. The horizontal arrows indicate the upper and the bottom interfaces of the Ba122:P film. From this image the thickness is determined to be 185 nm using the observed film layer thickness of 308 nm and the observation angle of 31° (185 nm = 308 nm × tan 31°). This procedure was applied to all the samples that we measured transport properties.

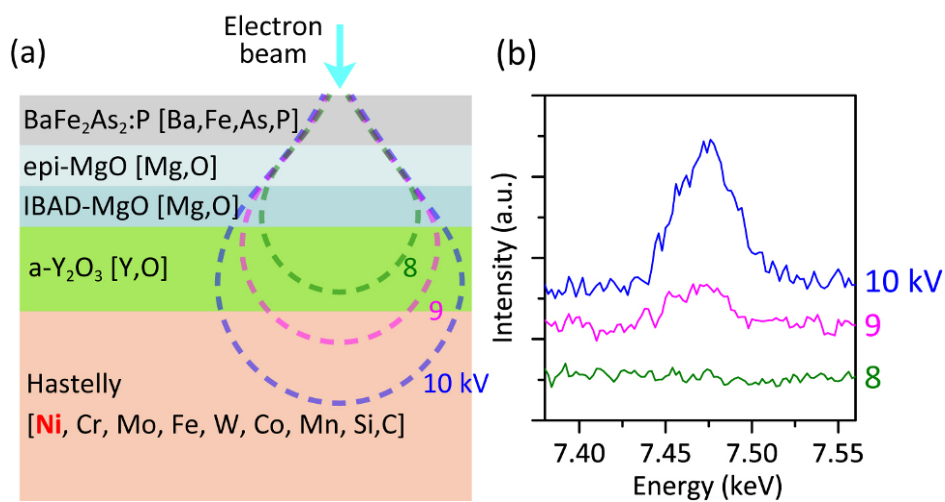

**Figure S2. Determination procedure of the acceleration voltage of the electron beam in the electron-probe microanalyzer.** (**a**) Schematic cross-sectional illustration of the sample and the penetration of electron at 8 – 10 kV. (**b**) The Ni Kα spectra at 8 – 10 kV. Based on these spectra, we performed the chemical composition analysis at 8 kV.